\documentclass[openacc]{rstransa}

\titlehead{Research}

\begin{document}

\title{Radiation doses during extreme solar energetic particle events}

\author{
S.~Dalla$^{1}$, K. Herbst$^{2}$, R. Muscheler$^{3}$ and M.J. Owens$^{4}$}

\address{$^{1}$Jeremiah Horrocks Institute, University of Lancashire, Preston , PR1 2HE, UK\\
  $^{2}$Centre for Planetary Habitability, Department of Geosciences, University of Olso, Norway\\
  $^{3}$Department of Geology, University of Lund, Sweden\\
$^{4}$Department of Metereology, University of Reading, UK
}

\subject{Astrophysics, Space exploration}


\corres{Silvia Dalla\\
  \email{sdalla@lancashire.ac.uk}}

\begin{abstract}
  Ions and electrons accelerated to high energies during flares and coronal mass ejections at the Sun may escape the solar atmosphere and, guided by the interplanetary magnetic fields, propagate through space to near-Earth locations. These Solar Energetic Particles (SEPs) can be detected directly by spacecraft instrumentation. The highest energy SEPs may also propagate through the geomagnetic field and precipitate to low atmospheric heights, producing secondary particles including neutrons and protons that trigger the formation of cosmogenic radionuclides. The space weather effects associated with the SEP ion population (for the most part protons) consist principally of radiation risk to aviation, humans in space and spacecraft. This paper focusses on the risks to aviation and astronauts and emphasizes how the parameters of the SEP event, including  fluence and spectrum, affect radiation doses. Space weather effects for events that took place in recent decades, for which a large body of measurements and models exist, are discussed.  SEP events of extreme magnitudes, such as those extrapolated from radionuclide data from the distant past are then considered and first estimates of the associated radiation at aviation altitudes and in space presented. These are worst-case estimates derived within the assumption that the radionuclide spike was caused by a single SEP event and that the particle spectrum and geomagnetic conditions were the same during all events.
\end{abstract}


  

\maketitle

\section{Introduction}

Solar Energetic Particles (SEPs) are ions and electrons accelerated during solar eruptions and able to escape into interplanetary space \cite{Kle2017}. In many cases the solar events that produce SEPs include  both a solar flare and a coronal mass ejection (CME), the latter often driving a shock through the corona and interplanetary space.
SEPs span a wide energy range, from 0.1 MeV up to about 10 GeV. Solar events capable to energise protons up to the GeV range are rare.
SEPs are thought to be accelerated during flare-associated magnetic reconnection and via shock acceleration at CME-driven shocks. A fraction of the high energy SEPs may propagate through the Earth's magnetic field and precipitate to low atmospheric heights where they produce a number of secondary particles via nuclear reactions, including neutrons and muons. As a result of the cascade of reactions that follow, among other atoms, cosmogenic radionuclides such as $^{14}$C, $^{10}$Be and $^{36}$Cl are formed via the  interaction  with atmospheric constituents.

SEP events are a key component of space weather since they pose a radiation risk to humans in space and at aviation altitudes \cite{Loc2007,Bai2023}. In addition, they may also cause single event effects, displacement damage and cumulative effects due to total ionising dose on spacecraft and aircraft electronics,  as well as internal charging. As a result SEPs can lead to damage to instrumentation, including in the worst cases even loss of satellites, and disrupt high-frequency  communications \cite{Can2013, Kra2016}. Thus a large effort in improving our understanding of the risk posed by SEP events, forecasting them \cite{Whi2023,Pap2025} and preparing mitigation strategies is ongoing.

The radiation effects of SEPs at or near Earth have been investigated for decades \cite{Xap2000,Tow2006,Jig2014,Fuj2021,Lar2024}. Over the years most assessments of radiation risk have been based on the assumption that the most dangerous events are those associated with Ground Level Enhancements (GLEs), measured as sudden increases in the neutron counts at ground level. Therefore GLEs, and in particular the most intense GLE in terms of neutron counts detected - GLE 5 on 23 February 1956 -  have been broadly considered as the worst case scenario in terms of space weather radiation and their impact assessed \cite{But2008,Blu2026}.

However recently it has been discovered that much more intense SEP events may have taken place in the distant past, based on high  resolution measurements of cosmogenic radionuclides in tree rings and ice cores \cite{Miy2012,Mek2015,Pal2022a,Bar2023}. Under the assumption that  the excess in the annual production rates of cosmogenic radionuclides over the galactic cosmic ray signal  results from short-lived SEP events, the particle fluences associated with the events have been estimated based on their enhancements \cite{Uso2023,Kol2023} and they appear in some cases to be about two orders of magnitude larger than those typical of GLEs.
In the following the large cosmogenic radionuclide  events will be referred to as extreme SEP events.

From a space weather perspective, it is important to assess the possible impacts of the newly discovered extreme SEP events. Only a few studies have addressed their radiation impacts so far.  Dyer at al (2018) have presented an analysis of GLE 5 and derived radiation doses at aviation altitudes and single event upset rates  \cite{Dye2018} . They discussed the cosmogenic radionuclide events of 774 AD and 994 AD and suggested that their impact could be estimated by scaling the values obtained for GLE 5 upwards by factors of 30 and 15 respectively, corresponding to their peak flux relative to that of GLE5. They concluded that radiation impacts for these events at aviation altitudes and close to the poles can be very significant and pointed out the need to develop relevant planning strategies. Mishev et al (2023) analysed radiation doses at aviation altitudes during the 774 AD event and studied their altitude dependence  \cite{Mis2023JSWSC}. They advocated for the 774 AD event to be used as worst-case scenario for planning purposes.

This paper describes the radiation effects of SEP events and the approaches that can be taken to quantify the associated radiation doses, with emphasis on applications to aviation and astronauts in space. It presents examples of how the relevant radiation parameters have been calculated from data and models of GLE events. It then provides an estimate of radiation doses associated with the extreme events deduced from radionuclide data  for the 5 largest cosmogenic radionuclide events known at the present time.

\section{Radiation doses associated with SEP events}

The radiation environment in space includes contributions from SEPs, which are transient and difficult to predict, and from galactic cosmic rays (GCRs), which are always present as contribution to radiation and are weakly modulated by the solar cycle \cite{Cha2021}.
The highest radiation risks are to astronauts in deep space since they are not protected by shielding from the Earth's magnetic field. Space exploration, with potential trips to Mars having been proposed, poses a big risk to astronauts due to the long duration of the required space travel (e.g. about 600 days of cruise time to Mars) \cite{Loc2007}. Astronauts at the International Space Station (ISS) are also subject to significant radiation risks.
Aviation crew and passengers may be exposed to radiation during large SEP events, especially for polar flights \cite{Dye2003}, although to a smaller degree compared with people in space, thanks to the shielding provided by Earth's magnetic field.  In addition single event effects (SEEs), localised disruptions to electronic components caused by hits by high energy particles, are of concern for aviation since they can affect systems vital to aircraft control, navigation and communications.

SEP ions are particle radiation capable of producing ionisation when passing through bodies and materials. This can give rise to damage to living cells and  their DNA \cite{Kie2005, Cha2021}. The effects on humans include so-called deterministic and stochastic effects. The former are acute effects that take place when radiation levels exceed a threshold: they give rise to radiation sickness, organ damage, cognitive impairment and may lead to death. The latter involve a raised long term risk of ill health including cancer and genetic mutations, where even low doses result in an increase in risk. Frameworks for the protection of workers exposed to radiation, for example astronauts and aviation crews, have been developed  \cite{Kie2005, Cha2021}. 

For SEP events, the most numerous ionic species consists of protons. Heavier ions are also present in much smaller numbers.
Nuclear reactions of SEP ions with atmospheric constituents (e.g.~N$_2$, O$_2$)  generate secondary particles, including neutrons, electrons and muons.
The interaction of protons, heavier ions and secondaries with matter deposits energy in the body or material traversed.
To describe how energy is transferred to the matter exposed to radiation the linear energy transfer (LET) is introduced: LET is the energy locally imparted per unit path length by ionising radiation. Indicating LET as $L$ one can write:
\begin{equation}
  L = - \frac{dE_k}{dx}
 \end{equation}
where $E_k$  is the kinetic energy of the particle losing energy and $x$ indicates the distance travelled \cite{Kie2005}. Broadly speaking $L$ is is directly proportional to the charge number $Z$ of the ion under study and inversely proportional to its energy. Thus as a particle propagates through a material, it slows down and its LET increases. At the end of a particle's range in a material there is a steep increase in energy deposition called a Bragg peak  \cite{Cha2021}. 

The radiation dose $D$ is defined  as the energy absorbed per unit mass and is measured in Grays (Gy).
$D$  is related to the LET via the equation:
\begin{equation}
  D = \frac{F \, L}{\rho}
 \end{equation}
where $F$ is the particle fluence (event integrated flux) and  $\rho$ is the density of the absorbing medium.
Therefore $D$ is directly proportional to the particle fluence.  Calculating total radiation doses requires summing contributions from all the components of the radiation, e.g. adding the contributions of primary particles and secondary particles such as neutrons.

When applying the analysis of radiation doses to the case of human bodies for the purpose of radiation protection, a number of empirical definitions are introduced.
The first is the concept of equivalent dose, which
takes into account  the fact that  different types of radiation cause different biological effects. The equivalent dose is obtained by multiplying $D$ by the weighting factor associated with the type of particle radiation under study \cite{Kie2005}. For example the weighting factor for protons is lower than that associated to $\alpha$-particles and heavier nuclei.

Secondly it is necessary to take into account the fact that the absorption of radiation varies depending on  which part of the human body (organs and tissues) is being irradiated.
This leads to the definition of the effective dose, obtained by multiplying the organ equivalent doses by the organ weighting factor, repeating over all organs and adding the resulting contributions \cite{Kie2005}.
Effective dose  is measured in Sievert (Sv).

The threshold between deterministic and stochastic radiation effects is not clear cut, however approximate threshold effective dose values can be used. A typical effective dose threshold value for deterministic effects is $E_t$=0.7 Sv \cite{Kie2005}, with the understanding that this exposure would take place over a fairly short period of time, e.g.~of the order of days.

From the point of view of radiation safety, the International Commission on Radiological Protection (ICRP) has put forward a limit of 1 mSv as the annual dose limit for members of the public \cite{Icr2007}. The annual limit for radiation workers is much higher at 20 mSv, averaged over five years, with no single year exceeding 50 mSv.
The ICRP thresholds relate to exposures other than  the normal background radiation experienced by people on the surface of the Earth. Background radiation is due to small levels of radionuclides
present in the subsurface, building materials and atmosphere and from cosmic ray
neutrons. The contribution to the radiation doses from these sources is  2 to 3 mSv per year, but can be a few times higher in regions where subsurface minerals contain higher levels of radionuclides and radon gas.
 Many countries have incorporated and
adapted the ICRP limits into laws that protect workers and the general public from
excessive radiation exposure.

\section{Radiation doses during GLE events}

The radiation implications of SEP events have been studied by a number of authors using variety of methodologies and assumptions \cite[e.g.]{Bai2023,Cop2019,Han2022,Lar2025}.
The calculation of doses at aviation altitudes is complicated by the fact that the precipitation of energetic protons to low heights in the magnetosphere is highly dependent on latitude, with high latitudes being more accessible than low latitudes. To describe this effect the concept of geomagnetic cutoff rigidity is used, where for a given location on Earth the cutoff rigidity indicates the lowest particle rigidity (momentum per unit charge) able to propagate to that location.
A model of the geomagnetic and magnetospheric  fields is required to determine cutoff rigidities.
The existence of cutoff rigidities means that given an SEP spectrum outside the magnetosphere, for most locations on Earth only the high energy portion of the spectrum will contribute to the radiation doses at low altitudes in the atmosphere or at the ground. The radiation at aviation altitudes is mostly due to neutrons generated by the interaction of the primary SEPs with the atmosphere.

Copeland and Atwell 2019 \cite{Cop2019} analysed the largest SEP event of the neutron monitor era, GLE 5, which took place on 23 February 1956,  and derived radiation doses at flight altitudes for the event.  Note that it is customary to quote aircraft altitude in feet, following standard international aviation practice.
They considered two models of the SEP fluence spectrum for the event and calculated the effective doses at different cutoff rigidities.
They found that the radiation dose associated with SEPs has a strong dependence on latitude, being several orders of magnitude larger at the geomagnetic pole compared to mid and low latitudes. Their work showed that at low latitude regions  the contribution from GCRs to effective doses is larger than that from SEPs. They also demonstrated how the specific shape of the spectrum influences the radiation dose.
Overall for GLE 5 they obtained a value of the effective dose at 35,000 ft (corresponding to 10,668 m) and latitude close to the magnetic pole that exceeds the 1 mSv level, the recommended annual upper threshold  for a member of the public, as defined by the ICRP \cite{Icr2007}.
It should be noted that in many cases aircraft for commercial long-distance passenger travel operate at altitudes above 35,000 ft, e.g.~around 41,000 ft. In some cases altitudes for private passenger travel can reach 51,000 ft. In these situations the radiation doses will be higher than the 1 mSv level, due to reduced atmospheric screening.

\begin{figure}[b]
\centering\includegraphics[width=0.7\textwidth]{dose_vs_fluence_plots_fig1.jpg}
\caption{Effective dose at typical aviation altitudes (35,000 feet) at the geomagnetic pole for the cosmogenic radionuclides events and for GLE 5. The shaded area indicates the region where the calculated dose exceeds the approximate threshold for acute effects, $E_t$ (dashed line). }
\label{fig_aviationaltitude}
\end{figure}


Larsen and Mishev (2025) \cite{Lar2025} studied the radiation impact of the so-called Halloween events, which took place during October and November 2003 and gave rise to three GLEs: GLE 65, 66 and 67. Their focus was on the calculation of the effective doses at a flight altitude of 35,000 ft. This was carried out by using a model of the disturbed geomagnetic conditions during each GLE event and using the OTSO model to calculate cutoff rigidities.
They found values of effective doses much smaller than those obtained in  \cite{Cop2019}  due in part to the lower fluence of GLEs 65, 66, 67 compared to GLE 5.

\section{Estimating radiation doses for extreme SEP events} \label{sec_extreme}

The large fluences of cosmogenic radionuclide events compared to the GLEs of the modern area are likely to result in much larger radiation doses compared to GLE events. Here we provide  first estimates of the effective doses associated to the extreme solar particle events.
Deriving accurate doses for events in the distant past is complicated by the fact that we have limited information on the SEP spectrum and on the geomagnetic field during the events. In addition there are uncertainties in the event amplitudes. In the calculation below it will be assumed  that the radionuclide peaks were produced in a single SEP event  comparable to recent large events, as opposed to being associated to multiple SEP events.

A first estimate of radiation doses associated with cosmogenic radionuclide events can be obtained by using a scaling based on the values obtained for GLE 5, i.e. by calculating the effective dose $E$ as:
\begin{equation}
E=\frac{F_{200} * E_{GLE5}}{ F_{200,GLE5}}
\end{equation}
where $E_{GLE5}$ is the effective dose for GLE 5 at the geomagnetic pole, equal to 6.1 mSv \cite{Cop2019}, 
$F_{200}$ is the fluence of the extreme event under study and  $F_{200,GLE5}$ that of GLE 5.
Fluences are total event fluences at proton energies $>$200 MeV as estimated by
Koldobskiy et al 2023 \cite{Kol2023}.
 We choose $F_{200}$ as representative fluence parameter because protons of energy greater than about 100 MeV  are capable of travelling to low altitudes in Earth's atmosphere, where they produce secondary particles that give rise to radiation at aviation altitudes. Using this parameter is standard practice in the study of comogenic radionuclides \cite{Kov2014}.

We consider the confirmed extreme SEP events of 994 CE, 775 CE, 660 BCE and 7176 BCE, and the candidate event of 12350 BCE \cite{Bar2023}. The fluence value for the   latter event was estimated by Gobulenko et al 2025 \cite{Gob2025}. The fluence value at $>$200 MeV for GLE 5 is from of Koldobskiy et al 2021 \cite{Kol2021}.

\begin{figure}[t]    
\centering\includegraphics[width=0.7\textwidth]{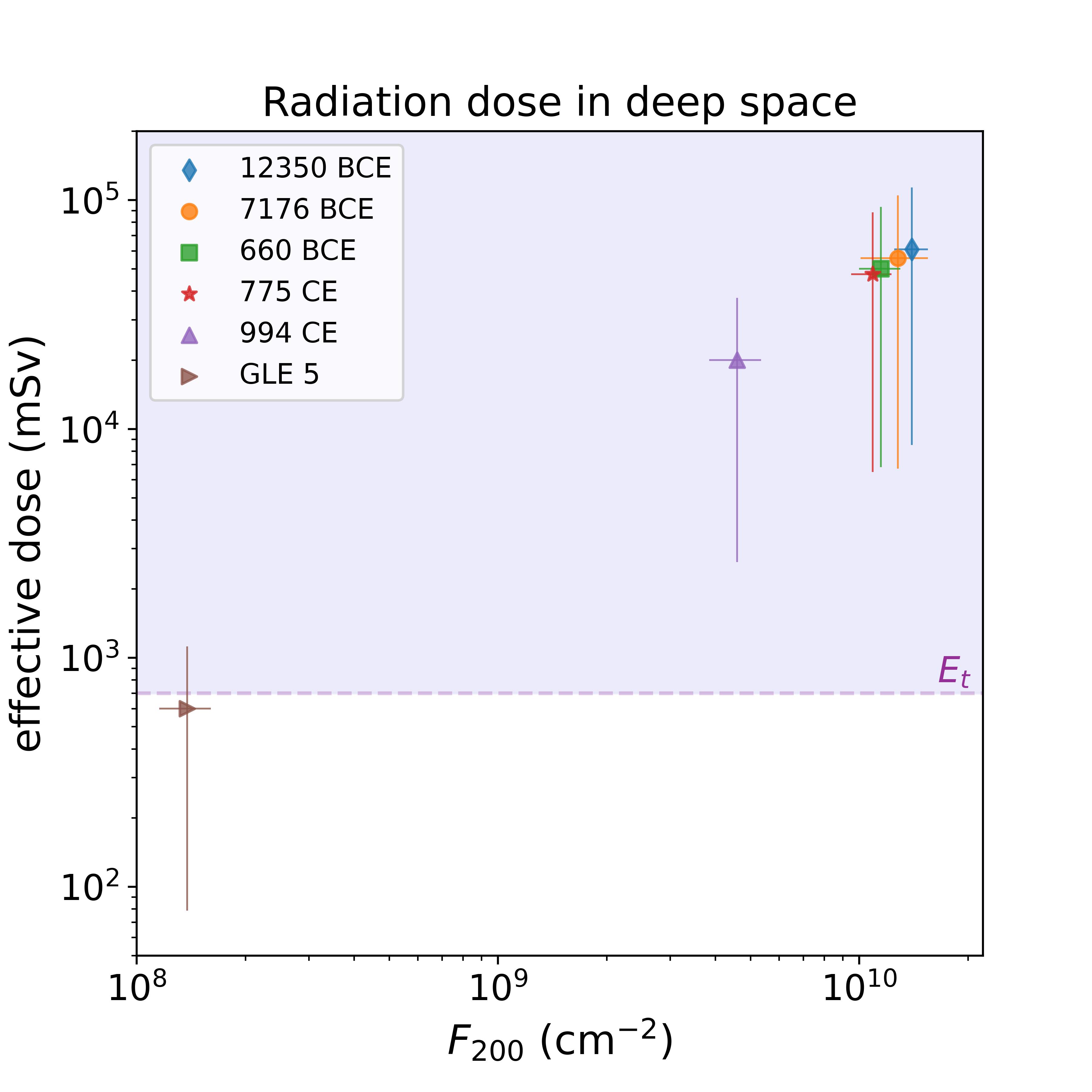}
\caption{Effective dose in deep space for the cosmogenic radionuclides events and for GLE 5. The shaded area indicates the region where the calculated dose exceeds the approximate threshold for acute effects, $E_t$ (dashed line). }
\label{fig_deepspace}
\end{figure}

Figure \ref{fig_aviationaltitude} shows the effective dose estimates at aviation altitudes
for the extreme SEP events and for GLE 5.  Errors were calculated using the published uncertainties in the values of $F_{200}$ \cite{Kol2023, Gob2025} and the fact that the error on the effective dose for GLE 5 at the geomagnetic pole was estimated to be 25\% \cite{Cop2019}. The uncertainty in the value of  $F_{200}$  value for GLE 5 is from \cite{Kol2021}.

The shaded region in  Figure \ref{fig_aviationaltitude}  indicates effective does above the approximate threshold for acute effects $E_t$=0.7 Sv. The plot shows that for the extreme events of highest fluences, the radiation doses at aviation altitudes over the pole are just below the threshold for acute effects.

Previous estimates of the route ambient dose equivalent at 40,000 ft and at 0 GV rigidity give a value of 210 mSv for the 774 AD event (obtained from a value of its peak flux as 30 times that of GLE 5) \cite{Dye2018}. This value is about half that shown in Figure \ref{fig_aviationaltitude} but of the same order of magnitude. For the event of 994 AD earlier estimates gave a dose of 105 mSv \cite{Dye2018}, while our value is 200 mSv.

The radiation risk for humans in deep space due to an SEP event  is considerably higher than that at aviation heights.
A precise calculation of the former requires information about the precise location in space and the type of shielding that an astronaut is protected by, as well as the spectral shape of the event. The type and thickness of shielding is important to determine which fraction of the SEP protons are able to pass through the material and the properties of the secondary particle populations, including neutrons, that they  generate.
A full calculation of the radiation doses in space, keeping the above factors into account, is beyond the scope of this paper. However we can obtain an
an indication of the risk by using results of previous work in which radiation doses have been calculated both at aviation altitudes and in interplanetary space for the same event.
Matthi{\"a}  et al (2018) \cite{Mat2018} analysed the 10 September 2017 event (GLE 72) and found that the event produced effective doses at aviation heights (40,000 ft) over the pole of around 25 $\mu$Sv, while for a person  in a spacecraft in interplanetary space, protected by shielding with thickness corresponding to 30 g/cm$^2$, the dose was around 8 mSv. Thus the dose in space was about 300 times larger for this specific event.

Using a conservative estimate that the dose in interplanetary space is 100 $\pm$ 80 times larger than at aviation heights, the data points of Figure  \ref{fig_aviationaltitude}  have been multiplied by this factor to obtain an estimate of effective doses in space. 

Figure \ref{fig_deepspace} shows the effective dose estimates in deep space
for the extreme SEP events and for GLE 5.
Since all cosmogenic radionuclide events shown are above the threshold $E_t$,
it is clear that the magnitude of extreme events is such that extremely adverse effects would be experienced by astronauts in deep space during this type of events.




\vspace*{-7pt}

\vspace*{-5pt}


\section{Conclusions}

Several cosmogenic radionuclide spikes thought to be associated with solar events have been discovered  in tree ring and ice core datasets and dated to times between 12,350 BCE and 994 CE. Particle fluences during these extreme solar events have been calculated to be significantly higher than in GLEs,  events during the last 100 years that are commonly regarded as the worst-case scenarios for the assessment of space weather radiation impact.
Currently cosmogenic radionuclide events are not included in radiation risk assessments for astronauts.
When assessing the risk from extreme SEP events is important to stress that they are rare events, with a frequency estimated to be about once every few thousand years.

In this paper we discussed  the  radiation implications of extreme solar energetic particle events.
We considered five cosmogenic radionuclide events and provided first estimates of the associated effective radiation doses based on simple scaling of the values of GLE 5.
These extrapolations indicate that the radiation dose from extreme solar particle events in deep space is likely to be above the threshold for deterministic effects, meaning that they could be fatal to astronauts.
For the largest events it is possible that the effective doses may be over  the threshold for direct acute radiation effects also at flight altitudes. The long term risk of ill health for people exposed to extreme events is likely to be high.

It should be noted that there are many uncertainties and approximations in the extrapolations: Figures \ref{fig_aviationaltitude} and \ref{fig_deepspace}  implicitly assume  that the same  geomagnetic configuration  and spectra were present during all events. It is very likely that this was not the case and future work will be needed to carry out a more detailed analysis.
There is also a possibility that the cosmogenic radionuclide events may be due to multiple solar particle events, rather than a single event. This would mean that the maximum fluence associated to a single SEP event may be significantly smaller than the estimates obtained. 
Should this be the case, if a  person were exposed to a single event, rather than the multiple event episode, the associated radiation dose would be significantly smaller than reported in Section \ref{sec_extreme}.
However the integrated effects over long timescales would remain the same, for example for an astronaut on a 2-year journey to Mars.

Our overall conclusion is that radiation effects associated with events of magnitude similar to the cosmogenic radionuclide events is likely to be very significant (above the threshold for acute radiation effects at many locations in space and potentially also at aviation altitudes) and that they need to be considered as part of future radiation risk assessments.

 




\ack{S.D.~acknowledges support from the UK Science and Technology Facilities Council (STFC) through grant No. ST/Y002725/1. This project further has received funding from the Research Council of Norway through the Centres of Excellence funding scheme, project number 332523 (PHAB).
  The research was partly funded by the European Union (ERC, PastSolarStorms, 101142677 to RM). Views and opinions expressed are however those of the author(s) only and do not necessarily reflect those of the European Union or the European Research Council. Neither the European Union nor the granting authority can be held responsible for them. MO is part-funded by STFC grant number UKRI1207 and Natural Environment Research Council (NERC) grant number NE/Y001052/1.}
\ack{Data Availability statement:
The data presented in the paper are available in the file radiation\_data.txt, available as supplementary material. The particle fluences F200 for GLE 5 and the radionuclide events were obtained from Koldobskiy et al 2021, 2023 and Gobulenko et al 2025.}


\bibliographystyle{RS}
\bibliography{radionuc_biblio}
%




\end{document}